# Atomically perfect torn graphene edges and their reversible reconstruction

Kwanpyo Kim[1,#], Sinisa Coh[1], C. Kisielowski[2], M. F. Crommie[1], Steven G. Louie[1], Marvin L. Cohen[1], and A. Zettl[1,*]

[1]*Department of Physics and Center of Integrated Nanomechanical Systems, University of California at Berkeley, and Materials Sciences Division, Lawrence Berkeley National Laboratory, Berkeley, CA 94720, U.S.A.*

[2]*National Center for Electron Microscopy, Lawrence Berkeley National Laboratory, Berkeley, CA 94720, U.S.A.*

[#]*Present address: Department of Chemical Engineering, Stanford University, Stanford, CA 94305, U.S.A.*

* To whom correspondence should be addressed: azettl@berkeley.edu

**The atomic structure of graphene edges is critical in determining the electrical, magnetic, and chemical properties of truncated graphene structures, notably nanoribbons. Unfortunately, graphene edges are typically far from ideal and suffer from atomic-scale defects, structural distortion, and unintended chemical functionalization, leading to unpredictable properties. Here we report that graphene edges fabricated by electron-beam-initiated mechanical rupture or tearing in high vacuum are clean and largely atomically perfect, oriented in either the armchair or zigzag direction. Via aberration-corrected transmission electron microscopy, we demonstrate reversible and extended pentagon-heptagon (5-7) reconstruction at zigzag edges, and explore experimentally and theoretically the dynamics of the transitions between configuration states. Good theoretical-**

**experimental agreement is found for the flipping rates between 5-7 and 6-6 zigzag edge states. Our study demonstrates that simple ripping is remarkably effective in producing atomically clean, ideal terminations thus providing a valuable tool for realizing atomically-tailored graphene and facilitating meaningful experimental study.**

**Introduction**

Manipulation of graphene edges at the atomic level is of fundamental importance in exploiting graphene's recognized potential in next generation electronic, optical, mechanical, and chemical devices[1-7]. For example, the electronic band-structure of graphene nanoribbons depends strongly not only on ribbon width but also on the detailed edge termination[1-4]. Zigzag (ZZ) and armchair (AC) graphene edges have distinct electronic states and scattering properties[1-6] as well as unique chemical properties[5-7]. Even though theoretical studies[1-7] have shed light on important aspects of bare and functionalized graphene edges, experimental observations and manipulation of "ideal" graphene edges at the atomic scale have been difficult to achieve, especially for suspended samples not influenced by substrate bonding and charging effects.

Available investigations of graphene edges have revealed that edges are prone to intrinsic and extrinsic modifications such as atomic-scale defects, structural distortions, and inhomogeneous and often unintended chemical functionalization. For example, most top-down fabrication processes including lithography, oxidative unzipping, and catalytic etching with metal result in highly defective edge structures[8-12]. Recently, anisotropic etching of graphene with hydrogen at elevated temperature has been used to produce nominally ZZ edges[13,14] but a direct atomic-scale characterization of the edge quality remains lacking. Bottom-up fabrication of graphene nanostructures has also yielded encouraging high-quality edge structure[15,16] but there are limitations in cleanly separating graphene from strongly-interacting metal substrates. Obtaining an atomically-precise and chirality-controlled graphene edge configuration is paramount to understanding truncated graphene's intrinsic properties and in enabling many graphene applications.

We have previously demonstrated that nicks in a tensioned suspended graphene membrane can be stimulated with an electron beam and thereby cause the membrane to catastrophically rupture or tear[17]. The direction of the tear (i.e. crack) follows almost exclusively the AC or ZZ direction, at least when viewed at the micrometer scale (the AC direction is more prone to tearing than the ZZ direction)[17]. However, the edge quality or configuration at the atomic scale has hitherto not been determined. Indeed, the alignment of graphene edges with high symmetry directions (AC or ZZ) at the micrometer scale does not guarantee perfect edge structure at the atomic level[12,18]; in principle, an edge that appears to be in the ZZ direction at the large scale could be composed of random (or collections of AC) edge structures at the atomic scale.

Previous experimental edge studies of graphene include micro Raman spectroscopy[14,15,18], scanning tunneling microscopy (STM)[15,16,19-21], and transmission electron microscopy (TEM)[12,17,22-26]. STM and TEM allow observations of graphene edges at the atomic scale including different electronic scattering properties and edge stability. Aberration-corrected TEM is an ideal tool for investigating graphene edges over relatively large areas with both atomic scale (sub-Å) spatial resolution and meaningful temporal resolution; it also eliminates unwanted substrate effects[12,17,22-26].

Here, we employ aberration-corrected TEM to demonstrate that graphene edges created by in-situ tearing of suspended monolayer graphene have clean, atomically-smooth edges in the ZZ or AC directions with over 90% edge configuration fidelity. We also observe extended pentagon-heptagon (5-7) reconstruction at the ZZ edge and demonstrate reversible transformation of the entire edge between different reconstructions. The atomic edge configurations are monitored in real-time and the edge-configuration-dependent dynamics and stability are analyzed, using both experiment and appropriate theoretical models. Effective activation barriers are extracted.

**Results**

**TEM imaging of torn graphene with AC and ZZ edges.** Fig. 1 shows atomic-resolution TEM images of a torn graphene edge nominally aligned with the AC lattice direction. The lower left side of Fig. 1a is a region of suspended single-layer graphene while the upper right side of the image shows vacuum. The inset to Fig. 1a is the Fourier transform of the image from which the overall lattice orientation is determined. The main image in Fig. 1a was taken with an accumulated electron beam dose less than $10^7$ e/nm$^2$ so as to capture the pristine edge configuration, as produced by the *in-situ* tearing process. (See Method section for the detailed procedure.) Notably, the torn graphene edge is extremely clean, regular, and straight even at the atomic scale. Fig. 1b shows a zoomed-in image near the edge, which reveals a perfect AC edge configuration (shown with atomic overlay in Fig. 1c). Figure 1d shows another segment of the graphene edge where a slightly irregular edge shape is revealed, with carbon atom vacancy defects. In this (rare) segment, four carbon atoms are missing from the perfect edge configuration. Overall, the AC torn edge segments typically exhibit fractionally perfect edge structure over 90% (See Supplementary Table S1 for detailed data). We emphasize that *in-situ* edge fabrication with electron-beam stimulation in a clean (high-vacuum) environment is key to obtaining atomically-smooth edge structures without functionalization.

Extended ZZ torn edges with atomically-smooth, ideal edge structure are also observed. Fig. 2a show an atomic-resolution TEM image of a torn edge aligned in the ZZ lattice direction. Fig. 2b is a zoomed-in image of the same graphene edge. Even though in these images it is non-trivial to resolve the location of each carbon atom due to sample tilting and electron-beam induced mechanical instability, we observe that there is a clear periodic intensity pattern at the graphene edge with a periodicity of around 4.9 Å, as marked with red arrows in Fig. 2b. This intensity pattern originates from a previously predicted[5,6] pentagon-heptagon (5-7) reconstruction at the ZZ edge (shown with atomic overlay in Fig. 2c). This reconstruction has been previously investigated via TEM over a limited range[22,24,27]. More clearly resolved atomic resolution images of 5-7 the reconstruction are presented later in this manuscript (see Figs. 3 and 4). The 5-7 reconstructed edge can be derived from the pure (6-6) ZZ edge with only local carbon bond rotations and has lower edge-energy than the (6-6) ZZ edge[5]. The experimentally obtained image of a 6-6 ZZ edge (i.e. without 5-7 reconstruction) is shown in Fig. 2d for comparison. The 6-6 ZZ edge shows an intensity pattern with the regular graphene lattice periodicity of 2.46 Å. Again, as demonstrated in Fig. 2 as well as in Fig. 1, tearing graphene is a highly effective method to obtain atomically clean, well-define graphene edges of specified chirality.

**Dynamics and stability of ZZ and AC edge configurations.** *In-situ* fabricated atomically-smooth ZZ and AC edge segments also provide excellent platforms for a detailed study of dynamics and stability of different edge configurations. Most previous TEM studies on graphene edges have relied on electron-beam sputtering onto the graphene lattice to produce edge configurations (such as the edge around a growing hole)[22,24,28]. During sputtering it is difficult to observe reversible transitions between bistable edge configuration states. In the present study extended edge configurations are readily available as pristine ZZ and AC edge configurations. We find that, under our electron illumination conditions, both the AC and ZZ edges show dynamical effects, with the ZZ edge being much more active, as we now describe in detail.

Fig. 3 shows a time series of TEM images of a relatively rare torn graphene edge corner. This corner area is chosen for monitoring since it shows flat ZZ and AC segments side-by-side, which facilitates a direct comparison of the dynamics. The electron-beam exposure and readout time are 0.5 second and 1.3 second per frame, respectively. In Figs. 3(a-e), five TEM images are shown where the sequential images are separated by 5 image frames. The up-sloping left-side and down-sloping right-side segments within each panel show the ZZ and AC edge configurations, respectively. Overall, both graphene edges are quite stable to e-beam induced sputtering at the experimental time scale (with beam current $2.1\ (\pm\ 0.1) \times 10^6$ e nm$^{-2}$s$^{-1}$).

Figs. 3(f-j) show the same sequential TEM images displayed with overlay edge representations. The red arrows indicate heptagon rings from 5-7 edge reconstructions. The observed periodicity of 4.92 Å in the ZZ region of Figs. 3f is in agreement with images shown in Figs. 2b,c which confirms that the ZZ tear edge of Fig. 2 has 5-7 reconstruction. The red dotted and blue solid lines represent 5-7 reconstructed and 6-6 ZZ edges, respectively. As clearly shown in the time-series of Fig. 3, under the influence of the electron beam the ZZ edge frequently undergoes dramatic, extended and fully reversible structural transitions between a 5-7 reconstructed edge and a 6-6 ZZ edge. In particular, Figs 3h,i show that the left-side ZZ segment can have a 100 % structure correlation with either 6-6 or 5-7 reconstruction. On the other hand,

the AC edge (shown with yellow dashed lines) is relatively stable under the electron beam, consistent with previous theoretical calculations[29]. Edge dynamics at the AC edge are mainly related to embedded short ZZ edge segments (which result in a one-unit dynamic "step" in the edge).

To examine ZZ reconstruction dynamics in more detail, the left-side, upsloping extended ZZ segments shown in Fig. 3 are monitored for flipping between 6-6 and reconstructed 5-7 ZZ configurations, presented in Fig. 4. We assign edge locations (from 1 to 16) in the ZZ segment as identified in Fig. 4a, and the edge configuration at each location is tabulated for a structural transition frame-by-frame. (See the Supplementary Movie 1 and Supplementary Table S2 for detailed data.) Fig. 4b shows the time evolution of the ZZ edge fraction during 89 time-frames. The rapid and frequent transformations between 6-6 and reconstructed 5-7 edge structures are clearly shown in data. For example, the 6-6 ZZ edge transforms to a 100% reconstructed 5-7 edge in two frames (2.6 seconds) during the time-frame 57-59. Green triangle data points in Fig. 4b represent other defect configurations such as adatom and vacancy defects. For the first 20 frames, part of the ZZ edge (edge locations 12 ~ 16) has an adatom (~ 0.2 in edge fraction) configuration and the edge starts to develop vacancy defects from frame 75 on. We find that the ZZ edge can effectively withstand knock-on damage for about ~ 50 time frames (65 seconds) in our experimental set-up. Overall, we observe more 5-7 reconstructed edge segments (66%) compared to 6-6 ZZ segments (32%) during frames 20-70. This is consistent with theoretical edge energy calculations which show that the 5-7 reconstructed edge has the lower energy (~ 1.7eV per a pair of hexagons) compared to the 6-6 ZZ edge[5,6].

**Discussion**

Interestingly and importantly, the transformation to/from 5-7 ZZ configurations is a collective behavior, with flips in one region highly correlated with nearby edge structure. We find that 5-7 reconstructions occur predominantly adjacent to preexisting 5-7 locations. To quantify this behavior, we first assign the 5-7 occupation value $a(x,t)$ for a pair of carbon rings at the edge, with $a(x,t)=1$ for a 5-7 ZZ edge pair and $a(x,t)=0$ for a 6-6 ZZ pair, respectively. Here $x$ (from 1 to 8) and $t$ (from 20 to 70) represent different locations of carbon-ring-pair and time frames. We then define a probability function $p$ for 5-7 reconstruction

$$p = \frac{\sum_{x,t} a(x,t)[a(x-1,t)+a(x+1,t)]}{2\sum_{x,t} a(x,t)} \qquad (1)$$

This probability function calculates, for a given 5-7 reconstruction edge site, the probability that a nearest neighbor site (left and right locations) is occupied by 5-7 edges. If the 5-7 reconstruction occupies random sites along the ZZ edge, we expect that $p$ is close to the average 5-7 edge fraction value, 66%. From our experimental data (See Supplementary Table S2), however, we obtain $p$ = 92%, which is significantly higher that the value with the random-location assumption. This demonstrates a high degree of correlation between reconstructed 5-7 edge sites. Theoretical calculations have shown that the activation barrier for the first 5-7 edge reconstruction is ~ 0.8 eV and can be lowered with the presence of a 5-7 reconstruction nearby[30]. This lowered activation barrier by nearby 5-7 reconstruction is consistent with the observed correlation of 5-7 edge sites.

We now consider the flipping rate at the ZZ edge between 6-6 carbon rings and 5-7 carbon rings, experimentally and theoretically. The activation barriers for the flipping are 0.8 eV (6-6 => 5-7) and 2.4 eV (5-7 => 6-6)[5,30], which are significantly higher than the thermal energy. Therefore, in our experiment, the high-energy incident electron-beam (80 keV) provides the energy for the transitions between 6-6 and 5-7 rings. The experimentally obtained flipping rates are 0.26 s$^{-1}$ (6-6 => 5-7) and 0.12 s$^{-1}$ (5-7 => 6-6), with a ratio R (6-6 => 5-7)/(5-7 => 6-6) ~ 2.3. Using the cross section for Coulomb scattering between an incident electron and a carbon atom[31,32], we can estimate the total cross section of scattering events where the energy above the threshold energy (0.8 eV or 2.4 eV) is transferred to a carbon atom. (See the Supplementary Note S1 for the detailed calculations) Under our experimental condition ($j = 2 \times 10^6$ e/s·nm$^2$), we obtain rates of 0.38 s$^{-1}$ (6-6 => 5-7) and 0.11 s$^{-1}$ (5-7 => 6-6), which gives a rate ratio R = 3.5 in good agreement with the experimental flipping rate ratio. (We assume that all the scattering events with energy transfers above the threshold energy result in a carbon ring flip process.) We find that thermal lattice vibrations[33] do not significantly change the atomic displacement rate and expected flipping rates. After taking the lattice vibrations into account, we obtain flipping rates of 0.44 s$^{-1}$ (6-6 => 5-7) and 0.13 s$^{-1}$ (5-7 => 6-6), with a ratio R ~ 3.4. (Please see the Supplementary Note 2 for the detailed calculations.)

Using the experimentally observed flipping rates, we can also estimate the effective activation barriers for 6-6 => 5-7 and 5-7 => 6-6 transformations. Fig. 4c shows the energy transfer rate to a single carbon atom as a function of transferred energy from the electron beam of 80 keV. We find certain effective cut-off energies, which reproduce experimentally obtained flipping rates (areas under the curve from the cut-off energy to maximum transfer energy). Using this procedure, we estimate that the effective activation energy barrier for the 6-6 => 5-7 flip is 1.1 eV, while for the 5-7 => 6-6 flip it is 2.3 eV (corresponding energy transfer rates are shown by shaded areas in Fig. 4c). When we take thermal lattice vibrations into consideration[33], the effective activation energy barriers increase by ~ 0.2 eV. (1.3 eV for 6-6 => 5-7 and 2.5 eV for 5-7 => 6-6). These effective activation energies are close to the calculated energy barriers required for these transformations (0.8 eV and 2.4 eV)[5,30]. One should note that here we have only considered the beam-induced displacement effect as the energy transfer mechanism and omitted other mechanisms. In fact, investigations of energy transfer mechanisms in electron microscopy are now actively pursued. They include ultrafast electron microscopy with ps of time resolution to addresses non equilibrium phonon excitations and subsequent long-wavelength atomic motion in thermalization processes[34], the role ionization processes[35], and heating effects[36]. The inclusion of other mechanisms in our analysis can result in larger values for the estimated activation energies. (See the Supplementary Note 3 for discussion on other types of possible energy transfer mechanisms.)

In conclusion, we demonstrate that torn graphene edges produced by e-beam-induced rupture along ZZ or AC directions are exceptionally clean and straight even at the atomic scale. The ZZ edge can be completely and reversibly flipped between two different metastable configurations, one with pure hexagons at the edge, the other with previously predicted 5-7 reconstructions. Flipping rates and activation energies are consistent with theoretical modeling. With the observed high energy barriers, we believe that the pure AC edge, and both of the ZZ-based edges, can be locked in and remain stable at room temperature.

**Method**

**Materials.** Graphene is obtained by chemical vapor deposition (CVD) on polycrystalline copper (99.8 % Alfa Aesar, Ward Hill, MA) with a growth temperature 1035 °C[37]. After synthesis, graphene is transferred to Quantifoil holey carbon TEM grids by a clean transfer process[17]. We use $Na_2S_2O_8$ solution to etch the copper substrate. The CVD graphene sample is mostly monolayer with the average grain size of above 5 μm[38]. The CVD graphene is suitable for preparing suspended graphene samples with large quantity, which enables us to systematically study in situ tearing of graphene edges.

**Atomic resolution TEM.** The atomic resolution TEM images of graphene edge were obtained with the TEAM 0.5 at the National Center for Electron Microscopy, Lawrence Berkeley National Laboratory. The microscope is equipped with image Cs aberration corrector and monochromator and was operated at 80 kV. The TEM image was taken at the over-focus 10 nm, which allows an optimal imaging condition with the bright atom contrast. In Situ Tearing of graphene and image acquisition were performed with vacuum pressures below $5\times10^{-8}$ Torr near the sample.

For single-shot TEM images (Figure 1 and 2), we went through the following steps to minimize the electron beam-induced damages to graphene torn edges. As shown in Supplementary Figure S1, we identify a preexisting tear on suspended graphene at low magnifications. Once we find an area of interest we temporarily block the electron beam. With a higher magnification, we set a focus and proper imaging setting on a sample area far away from the identified tear region. Then we move to a region, where we expect to find an in situ fabricated torn edge as shown in Supplementary Figure S1a. The extended tear is usually prone to mechanical instability such as vibration, which prevents atomic resolution imaging of torn edge. The graphene tear around an edge of carbon support generally has better mechanical stability and allows atomic-resolution imaging.

**Figures**

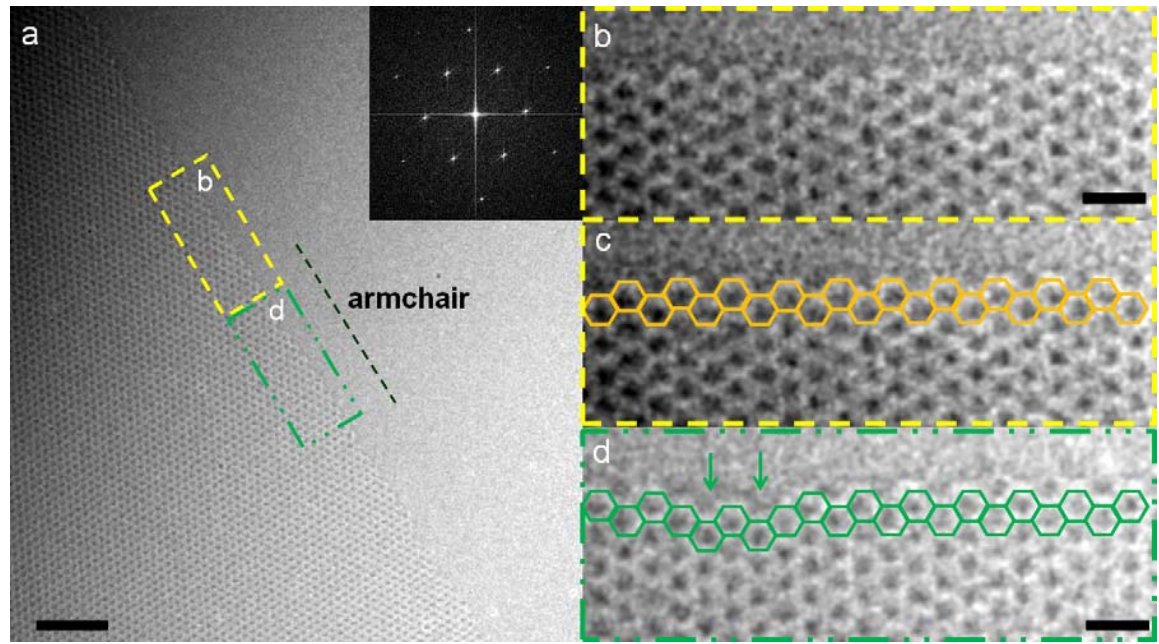

**Figure 1. Straight graphene torn edge with armchair (AC) edge configuration.** (a) Atomic resolution TEM image of a torn edge with AC configuration. The inset is the Fourier transform of the image. The yellow and green dashed boxes are the field of view for Figure b and d, respectively. Scale bar, 2 nm. (b) The zoom-in image of the graphene edge showing a perfect AC edge. Scale bar, 0.5 nm. (c) The same TEM image with atomic structure overlay. (d) The zoom-in image of the graphene edge at a different location showing a segment with local irregularity. The arrows indicate the locations of four missing atoms. Scale bar, 0.5 nm.

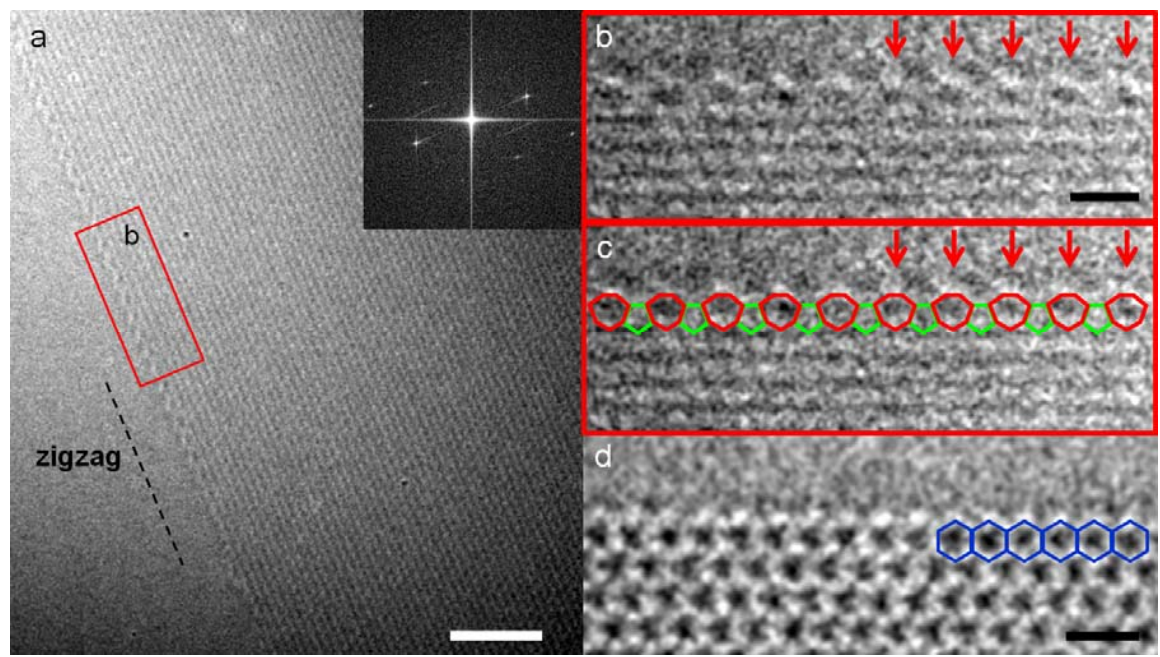

**Figure 2. Straight graphene torn edge with zigzag (ZZ) edge configuration.** (a) Atomic resolution TEM image of a torn edge in ZZ direction. The inset is the Fourier transform of the image. The red box is the field of view for Figure b. Scale bar, 2 nm. (b) Zoom-in image of the graphene edge. The red arrows indicate heptagon rings with 4.92 Å inter-ring distance. Scale bar, 0.5 nm. (c) The same figure with atomic overlay. The graphene edge shows a pentagon-heptagon (5-7) reconstructed ZZ edge configuration. (d) Graphene torn edge with pure (6-6) ZZ edge configuration (without reconstruction). The atomic edge configuration is overlaid at the right side of the image. Scale bar, 0.5 nm.

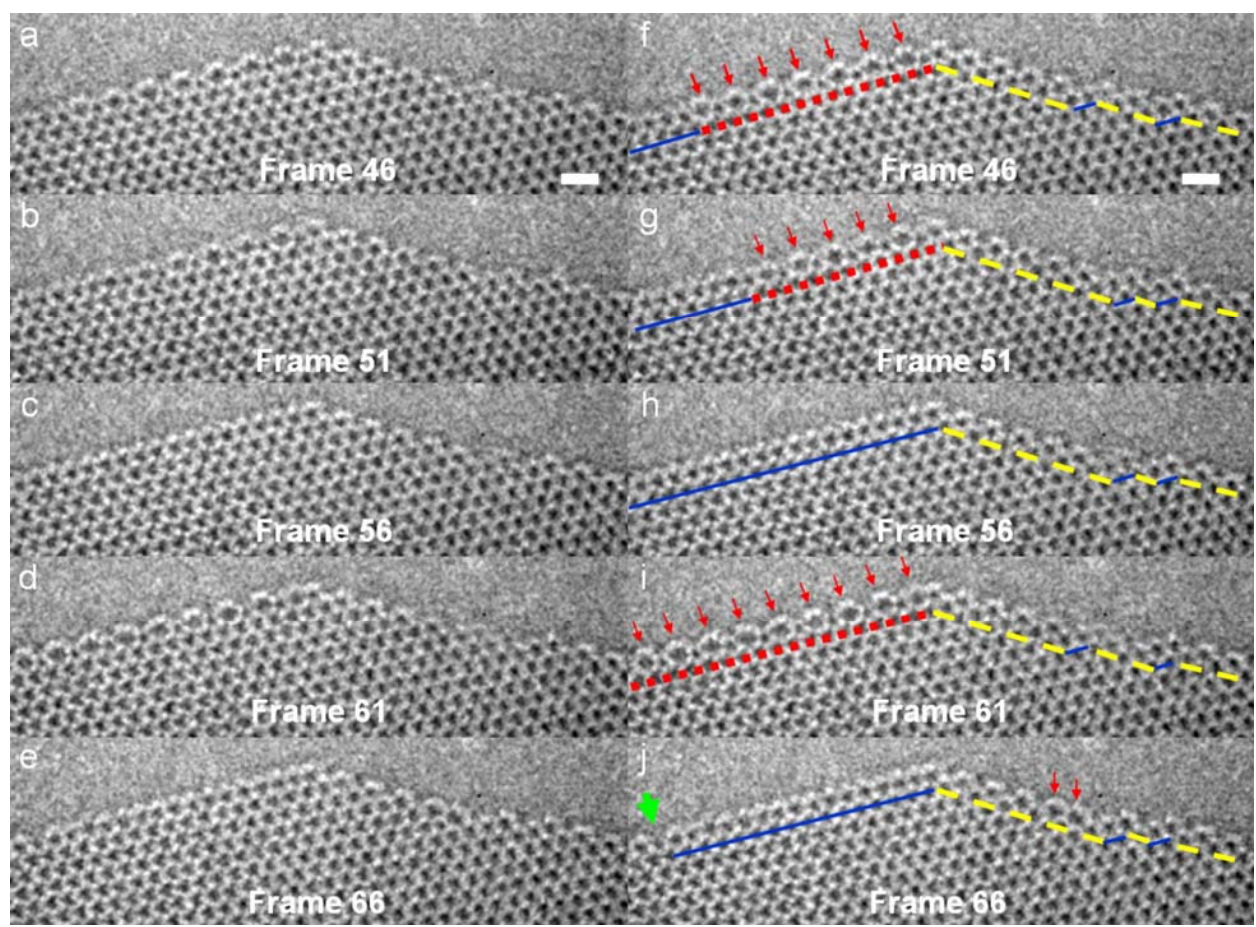

**Figure 3. A time-series of TEM images of a graphene torn edge under electron beam.** (a-e) A time-series of TEM images of graphene edge. Each image is apart from each other by five frames. The left (right) segments have the ZZ (AC) edge configuration. Scale bar, 0.5 nm. (f-j) The same sequential TEM images with edge representations. The red arrow indicates a heptagon ring. The blue solid and red dotted lines represent 6-6 ZZ and 5-7 reconstructed ZZ edges, respectively. The yellow dashed lines show AC edge configuration. The green arrow in Figure j shows a vacancy defect.

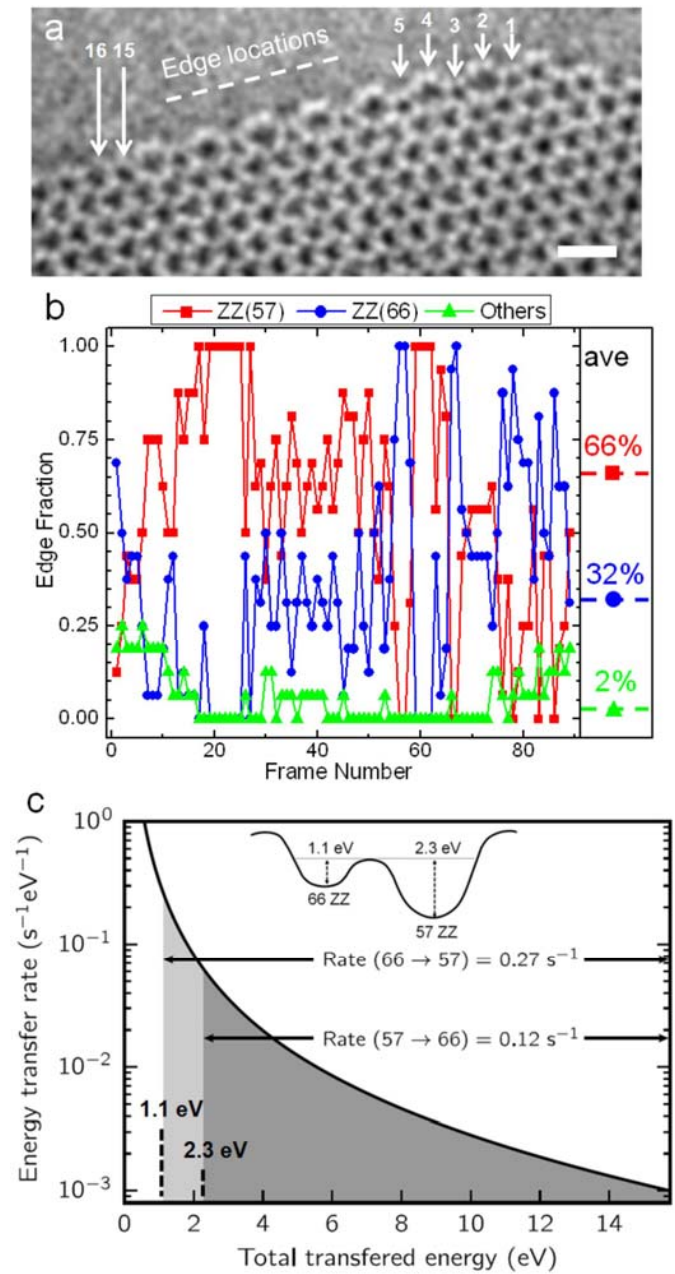

**Figure 4. Structural transitions between 6-6 and reconstructed 5-7 ZZ edge.** (a) TEM image of graphene ZZ edge with assigned hexagon locations. The edge locations from 1 to 16 are monitored for structural transition frame-by-frame. Scale bar, 0.5 nm. (b) Time evolution of ZZ edge fraction during 89 time frames. The average occupied edge fraction (with frames from 20 to 70) is also shown. (c) Energy transfer rate to a single carbon atom as a function of transferred energy from electron beam. The total energy transfer rates shown by shaded areas give the effective activation barriers for flipping events between 5-7 and 6-6 ZZ edges. Inset shows the energy landscape of different ZZ edges.

**Acknowledgement**

This research was supported in part by the Director, Office of Energy Research, Materials Sciences and Engineering Division, of the US Department of Energy under contract No. DE-AC02-05CH11231 which provided for TEM characterization, including that performed at the National Center for Electron Microscopy, and theoretical modeling; by the Office of Naval Research under MURI grant N00014-09-1066 which provided for graphene synthesis and suspension; and by the National Science Foundation within the Center of Integrated Nanomechanical Systems, under Grant EEC-0832819, which provided for additional sample characterization and personnel support.

**Author contributions**

K.K. and A.Z. designed the experiments. K.K. carried out the experiments. K.K. and A.Z. co-wrote the paper. All authors were involved in data analysis and commented on the manuscript.

**Additional Information**

**Supplementary Information** accompanies this paper at http://www.nature.com/naturecommunications

**Competing financial interests:** The authors declare no competing financial interests.

**Reprints and permission** information is available online at

http://npg.nature.com/reprintsandpermissions/

**How to cite this article:** Kim, K. et al. Atomically perfect torn graphene edges and their reversible reconstruction. *Nat. Commun.* 4:2723 doi: 10.1038/ncomms3723 (2013).

# Supplementary Information

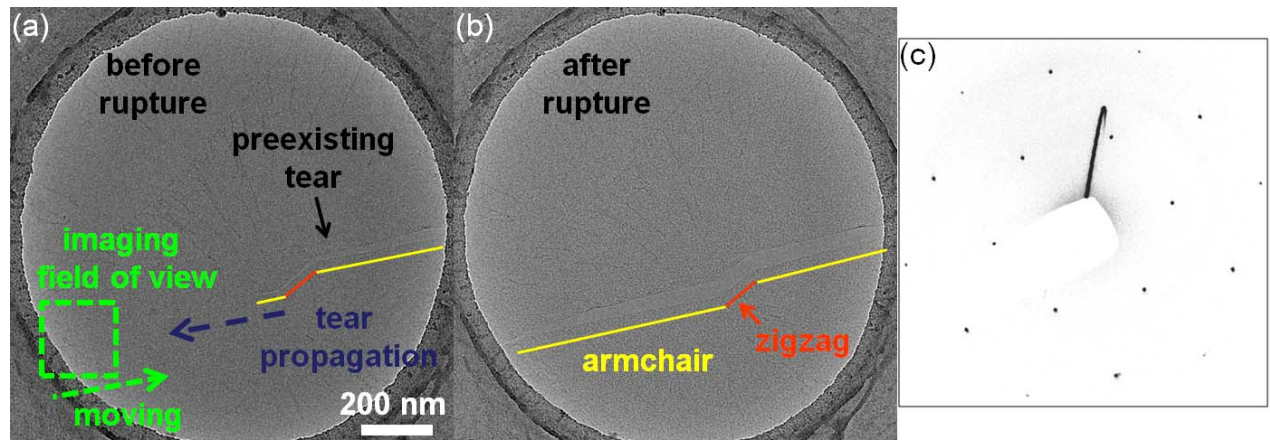

**Supplementary Figure S1. The detailed procedure for TEM imaging of graphene torn edge.** (a) TEM image of a graphene torn edge before the tear propagation. Once a tear is identified at low magnification, the electron beam is blocked to minimize the e-beam-induced damage. With a high magnification, we set a proper focus setting at a location far away from the identified tear. We guess its propagation direction with the shape of a graphene tear and start to image a sample approaching from the carbon support edge as shown. (b) TEM image of the same graphene torn edge after tear propagation with electron beam. (c) Diffraction pattern of the graphene membrane around the graphene edge. Microscopic edge directions can be assigned from the diffraction pattern.

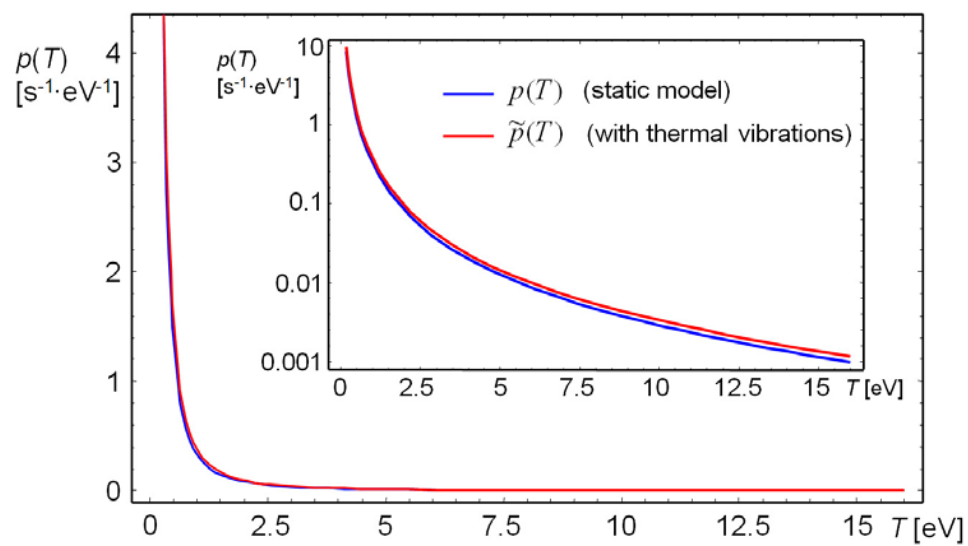

**Supplementary Figure S2. The calculated number of scattering events per second with transferred energy *T* shown as a function of *T* for single carbon atom.** Two plot show *p(T)* with and without the thermal lattice vibrations, respectively. The same *p(T)* are shown as the semi-log scale in the inset.

**Supplementary Table 1. The analyzed data set to study graphene edges in armchair (AC) and zigzag (ZZ) edge configurations.**

| Edge type | # of analyzed images | Edge length | Pristine edge length | Other edge length |
|---|---|---|---|---|
| Armchair (AC) | 6 | 90.2 nm | 82.1 nm (91 %) | 8.1 nm (9 %) |
| Zigzag (ZZ) | 4 | 43.6 nm | 39.1 nm (90 %) | 4.5 nm (10 %) |

**Supplementary Table 2. Detailed edge configuration (for location 1-16) of a zigzag edge analyzed in Figure 4.** Pentagon-heptagon (5-7) reconstructed edge segments are shown in red, while hexagon (6) rings are shown in blue. The raw data can be found in the Supplementary Movie.

5 = pentagon
6 = hexagon
7 = heptagon
a = adatom
v = vacancy

**Supplementary Note 1 - The cross section for Coulomb scattering between an electron and a carbon atom**

An analytic approximation of the cross section for Coulomb scattering between an incident electron and a nucleus[31,32] was employed to address the energy transfer rate to carbon atoms at the graphene edge during experimental imaging conditions. The scattering cross section for the events when energy $T$ or higher is transferred can be written as

$$\sigma(T) = \frac{4Z^2 E_R^2}{m_e^2 c^4}\left(\frac{T_{max}}{T}\right)\pi a_0^2 \left(\frac{1-\beta^2}{\beta^4}\right)\left\{1 + 2\pi\alpha\beta\sqrt{\frac{T}{T_{max}}} - \frac{T}{T_{max}}\left[1 + 2\pi\alpha\beta + (\beta^2 + \pi\alpha\beta)\ln\left(\frac{T_{max}}{T}\right)\right]\right\} \quad (S1)$$

where $Z$ is the atomic number of the target atoms, $E_R$ the Rydberg energy (13.6 eV), $a_0$ the Bohr radius of the hydrogen atom (5.3 × 10$^{-11}$ m), $\beta = v/c$ (electron velocity divided by the speed of light c), and $\alpha = Z/137$. $T_{max}$ is the maximum transferred energy in the scattering event and can be written as

$$T_{max} = \frac{2E(E + 2m_e c^2)}{Mc^2} \quad (S2)$$

under the assumption that the target atom mass $M$ is much heavier that the electron mass $m_e$ ($E$ is the kinetic energy of the electron). At 80 kV TEM operation, $T_{max}$ is 15.8 eV for carbon atoms. With the experimental imaging condition of electron beam intensity $j = 2 \times 10^6$ e/s·nm$^2$, we can calculate the number of scattering events per second as a function of the energy transfer $T$ to single carbon atom.

$$p(T) = -\frac{d\sigma(T)}{dT} j \quad (S3)$$

**Supplementary Note 2 - The effect of thermal lattice vibrations on the cross section for Coulomb scattering**

Recently, the thermal lattice vibrations were found to have an effect on the displacement cross section[33]. When the target carbon atom is not at rest due to the vibrations of the lattice, the transferred energy $T$ from incident electron can change. The modified maximum transferred energy can be written as

$$\tilde{T}_{max}(E_n) = \frac{(rc+t)^2}{2Mc^2} \tag{S4}$$

with $r = \frac{1}{c}\left(\sqrt{E(E+2m_ec^2)} + \sqrt{2m_ec^2 E_n}\right)$ and $t = \sqrt{(E+E_n)(E+2m_ec^2+E_n)}$

where $E_n = Mv^2/2$ is the initial kinetic energy of the target atom[33].

The mean square velocity of a carbon atom in graphene can be calculated as

$$\overline{v^2} = \frac{9k_b}{8M}\theta_D + \frac{9k_bT}{M}\left(\frac{T}{\theta_D}\right)^3 \int_0^{\theta_D/T} \frac{x^3}{\exp(x)-1}dx \tag{S5}$$

where $k_b$ is the Boltzmann constant, $T$ the temperature, and $\theta_D$ the Debye temperature (1287 K for graphene[33]). At room temperature $T = 293$ K, we found that the average kinetic energy of the target atoms, $\overline{E_n} = M\overline{v^2}/2 = 0.069$ eV.

To estimate the average maximum transferred energy $\tilde{T}_{max}$, we put the average kinetic energy $\overline{E_n}$ into the Eq. (S4), which gives $\tilde{T}_{max}(\overline{E_n}) = 17.9$ eV. Therefore, $\tilde{\sigma}(T)$, Eq. (S1) with modified maximum transferred energy $\tilde{T}_{max}(\overline{E_n})$, gives the scattering cross section for the events when energy $T$ or higher is transferred when we take thermal lattice vibrations into consideration. From this, the calculate the number of scattering events per second as a function of the energy transfer $T$ to single carbon atom is written as

$$\tilde{p}(T) = -\frac{d\tilde{\sigma}(T)}{dT}j \tag{S6}$$

The supplementary Figure S2 shows the calculated $p(T)$ and $\tilde{p}(T)$. We found that the effect of thermal lattice vibrations is not significant. The Figure 4c shows the $p(T)$ in the energy range to 15.8 eV.

**Supplementary Note 3 - Other radiation effects on graphene from high-energy electron**

We discuss different radiation effects on graphene from incident high-energy electrons. The important primary radiation effects[32,39] includes

1) electronic excitation or ionization of individual atoms,
2) collective electronic excitations (plasmons)
3) generation of phonons, leading to heating of the targets,
4) displacement of atoms (including sputtering of atoms)

The first effect, electronic excitation and ionization of individual atoms, is quickly quenched due to the high density of delocalized electrons in metal and graphitic materials, including graphene. The energy will be quickly dissipated throughout the specimen, preventing a direct ionization-induced ionic movement. Phonon can be generated by electron scattering with carbon nuclei but it is mainly generated from the dissipation of plasmons into phonon modes[32,39]. The energy transferred through processes from 1) to 3) together is related to internal thermal energy increase due to high-energy electron.

We can estimate the energy transfer rate to a specimen which is converted into internal thermal energy. If the beam-induced excitation of phonons is high enough, they will have an implication in the observed flipping rates between 57 and 66 zigzag edge configurations. Previously, for carbon samples, the average energy transfer value by one incident electron per unit mass thickness ($\Delta Q/\Delta x$) to specimen heating was found to be ~ 3 eVcm$^2$/μg[39]. Temperature increase of the specimen can be written as

$$\Delta T = \frac{j\rho}{4e\kappa} \frac{\Delta Q}{\Delta x} R^2 \tag{S7}$$

where $j$ is electron beam intensity, $\rho$ is the mass density of graphene (2.2 g cm$^{-3}$), $\kappa$ is the thermal conductivity of graphene (~ 4000 W m$^{-1}$K$^{-1}$)[40], $R$ is the distance between the center of electron beam to the location where the temperature stay constant (assumed to be 20 μm)[39]. From this, we obtain $\Delta T$ ~ 5 K with our experiment parameters. We find that the temperature increase due to inelastic scattering is not significant in our experiment set-up. Therefore, our assumption in the main manuscript that the observed transformation between 5-7 and 6-6 ZZ edge configurations is mainly the consequence of the displacement effects (electron − carbon atom scattering) is reasonable.

**Supplementary References**